\documentclass[conference, letter]{IEEEtran}
\usepackage{cite}
\ifCLASSINFOpdf
   \usepackage[pdftex]{graphicx}
\else
  \usepackage[dvips]{graphicx}
  \DeclareGraphicsExtensions{.eps}
\fi
\usepackage[cmex10]{amsmath}
\usepackage{tikz}
\usetikzlibrary{matrix,shapes,arrows,positioning,chains}

\usepackage{amssymb}

\usepackage{color}
\usepackage{algorithm,algpseudocode}
\usepackage{caption}

\usepackage{varwidth}
\usepackage{fancybox,framed}
\usepackage{flushend}
\usepackage{fixltx2e}
\usepackage{stfloats}
\makeatletter

\newcommand{\Rmnum}[1]{\expandafter\@slowromancap\romannumeral #1@}
\makeatother

\newcounter{lemmano}
\newcounter{propono}

\newcommand\copyrighttext{%
  \footnotesize \textcopyright 2021 IEEE}
\newcommand\copyrightnotice{%
\begin{tikzpicture}[remember picture,overlay]
\node[anchor=south,yshift=10pt] at (current page.south) {\parbox{\dimexpr\textwidth-\fboxsep-\fboxrule\relax}{\copyrighttext}};
\end{tikzpicture}%
}
\begin{document}
%
% paper title
% can use linebreaks \\ within to get better formatting as desired
% Do not put math or special symbols in the title.
\title{Quantum Computing for Artificial Intelligence Based Mobile Network Optimization}
%\author{\IEEEauthorblockN{\textsuperscript{} Jessica Moysen}
%\IEEEauthorblockA{
%\textit{Elisa Corporation, Helsinki, Finland }\\
%and Fundaci\'{o} i2CAT \\
%jessica.moysen@i2cat.net}
%\and
\author{
\IEEEauthorblockN{\textsuperscript{} Furqan Ahmed}
\IEEEauthorblockA{ 
\textit{Elisa Corporation}\\
Helsinki, Finland \\
furqan.ahmed@elisa.fi}
%\and
%\IEEEauthorblockN{\textsuperscript{} Mario Garc\'{i}a-Lozano}
%\IEEEauthorblockA{
%\textit{Universitat Polit\`{e}cnica de Catalunya}\\
%Barcelona, Spain \\
%mariogarcia@tsc.upc.edu}

\and
\IEEEauthorblockN{\textsuperscript{} Petri M\"{a}h\"{o}nen}
\IEEEauthorblockA{
\textit{Institute for Networked Systems} \\ 
\textit{RWTH Aachen University}\\
Aachen, Germany \\
pma@inets.rwth-aachen.de}

}

% use for special paper notices
%\IEEEspecialpapernotice{}

% make the title area
\maketitle
\copyrightnotice
% As a general rule, do not put math, special symbols or citations
% in the abstract or keywords.
\begin{abstract}
%A compelling approach to optimize certain discrete valued radio access network~(RAN) configuration parameters in an automated manner, is to apply combinatorial optimization formulation in conjunction with artificial intelligence (AI) approaches such as constraint satisfaction algorithms. Interestingly, many of the underlying problems are NP-hard, which makes them viable for exploring the potential and challenges of a quantum computing enabled RAN automation framework. 
In this paper, we discuss how certain radio access network optimization problems can be modelled using the concept of constraint satisfaction problems in artificial intelligence, and solved at scale using a quantum computer. As a case study, we discuss root sequence index~(RSI) assignment problem --- an important LTE/NR physical random access channel configuration related automation use-case. We formulate RSI assignment as quadratic unconstrained binary optimization~(QUBO) problem constructed using data ingested from a commercial mobile network, and solve it using a cloud-based commercially available quantum computing platform. Results show that quantum annealing solver can successfully assign conflict-free RSIs. Comparison with well-known heuristics reveals that some classic algorithms are even more effective in terms of solution quality and computation time. The non-quantum advantage is due to the fact that current implementation is a semi-quantum proof-of-concept algorithm. Also, the results depend on the type of quantum computer used. Nevertheless, the proposed framework is highly flexible and holds tremendous potential for harnessing the power of quantum computing in mobile network automation. %We conclude that quantum annealing technology, while highly promising for the future of AI enabled network automation, currently has limitations when it comes to commercial networks.
\end{abstract}

% Note that keywords are not normally used for peerreview papers.
%\begin{IEEEkeywords}
%Self-organizing networks, random access channel, root sequence index, graph colouring, greedy %algorithms.
%\end{IEEEkeywords}
% For peer review papers, you can put extra information on the cover
% page as needed:
% \ifCLASSOPTIONpeerreview
% \begin{center} \bfseries EDICS Category: 3-BBND \end{center}
% \fi
%
% For peerreview papers, this IEEEtran command inserts a page break and
% creates the second title. It will be ignored for other modes.
\IEEEpeerreviewmaketitle

\section{Introduction}
% The very first letter is a 2 line initial drop letter followed
% by the rest of the first word in caps.
% 
% form to use if the first word consists of a single letter:
% \IEEEPARstart{A}{demo} file is ....
% 
% form to use if you need the single drop letter followed by
% normal text (unknown if ever used by IEEE):
% \IEEEPARstart{A}{}demo file is ....
% 
% Some journals put the first two words in caps: is leading to emergence of a myriad of
% \IEEEPARstart{T}{his demo} file is ....
% 
% Here we have the typical use of a "T" for an initial drop letter
% and "HIS" in caps to complete the first word.
\IEEEPARstart{R}ecent years have witnessed a tremendous increase in interest towards quantum computing from the standpoint of both academia and industry. Several instances of phenomenal performance improvement over classical algorithms has garnered %critical acclaim
interest of scientific community and widespread media attention. Availability of cloud based quantum computing platforms has spurred a new wave of research and development activities leading to innovative solutions in different areas including traffic control~\cite{neukart2017traffic}, portfolio optimization~\cite{venturelli2019reverse}, finance~\cite{ORUS2019100028}, and machine learning in general~\cite{Biamonte2017}. However, the potential applications of this highly promising technology has not yet been investigated in the field of mobile networks, which are getting increasingly complex, and thus require more powerful optimization methods and computing platforms. Apart from~\cite{qml_6g}, existing works mostly focus on the applications of quantum computing from a perspective of security~\cite{MITCHELL2020101825}, quantum communication, or information theoretic aspects. In particular, contributions on upper layers and system level are minimal. 
In mobile networks, s{elf}-organization networking~(SON) algorithms based on classic computing approaches are often used for optimizing radio configuration management parameters.
Motivation for more intelligent automation such as zero-touch networks particularly stems from ever-increasing network densification, data rates, and heterogeneity. These challenges call for intelligent approaches capable of optimizing massive number of configuration management parameters on-the-fly\cite{challengesSON}. The optimization of these parameters constitute highly complex computational problems. Therefore, the performance of mobile networks is dependent on the computational intelligence and computing capabilities of the automation platform. To this end, use of artificial intelligence~(AI) in conjunction with quantum computing is a very promising research area, especially from a perspective of 5G and beyond. In fact, the computational problems underlying a number of network automation use-cases are purely discrete in nature and can be modeled as combinatorial optimization problems, where the objective is to find a correct values of parameters. Remarkably, a number of combinatorial optimization problems, especially the constraint satisfaction problems~(CSPs) can be tackled using a unified model of unconstrained quadratic binary programming~\cite{kochenberger2004unified}. Resulting approach can benefit immensely from advancements in scientific computing paradigms such as quantum computing, currently envisioned for the post-Moore's era~\cite{Hamiltonpostmoore}.
%is used to tackle a number of problems in the field of AI. 
In this paper, we discuss how radio access network (RAN) optimization use-cases can be formulated as CSPs, which paves the way for quadratic unconstrained binary optimization~(QUBO) formulation used in quantum computing. To this end, a generic architecture for quantum computing enabled mobile network automation is presented. We discuss physical random access channel~(PRACH) root sequence index~(RSI) assignment problem as an example use-case and compare the performance of classic and quantum computing approaches, on a CSP constructed using data from a commercial mobile network.

The rest of the paper is organized as follows: Section~\Rmnum{2}
discusses the CSPs and discusses how these can be applied to RAN optimization problems followed by a discussion on the preliminaries of quantum computing and proposes an architecture for quantum computing enabled mobile network automation platform. In Section~\Rmnum{3}, the proposed framework is applied or PRACH RSI parameter assignment, which is a use-case relevant to both 4G and 5G. Numerical results comparing classic heuristics and quantum computing approaches are discussed. Finally, conclusions are given in Section~\Rmnum{4}.

\section{Quantum Computing Enabled Framework}
\subsection{CSPs}
A generalized CSP is defined by a set of $n$ variables $x_1, x_2, \dots , x_n$, that take value from finite and discrete domains~$X_1, X_2, \dots , X_n$ and a set of constraints on their values. A constraint involves a subset of variables and defines allowed combination of values for that particular subset. Thus, each constraint is defined by a predicate, which is true iff the value assignment of the variables satisfies the constraint.  A state of the problem is defined by the assignment of values to the set or a subset of variables. The solution of CSP is an assignment of values to the set of variables that meets all constraints. The set of all possible assignments~$X_1 \times X_2 \times \dots  X_n$, is often called the solution space, as the solution is searched within this space. 
%\subsection{Applications in RAN Optimization}
A number of RAN parameter configuration problems can be formulated as CSPs. The constraints usually arise from the fact that the configuration of a cell is dependent on the network topology and needs to align with the configuration in neighboring cells. For instance, LTE/NR parameter PCI needs to be configured collision and confusion free. Collision free means that neighboring cells have different PCIs, whereas confusion free refers to the constraint that no two neighboring cells have the same PCI. Likewise, PRACH related parameters such as RSI and cyclic shift, carrier selection, and tracking area code (TAC) related problems can be modeled as CSPs.

Approaches for solving CSPs range from simple metaheuristics chosen according to the structure of underlying computational problem, to general constraint programming algorithms based on constraint activation and propagation principles. A comparison of local search with generic CSP algorithms for PCI is discussed in~\cite{LTE_JECE}. Metaheuristics aim at fast and strategic exploration of the search space of the solution, whereas constraint propagation rely on efficient and systematic search. Next, we discuss how to use quantum computing platform to solve these problems.

\subsection{Quantum Computing}
Theoretical foundations of quantum computing can be traced back to 1980s, when quantum mechanics and information theory, two highly influential areas, were unified giving rise to the field of quantum information theory. The use of quantum mechanical principles for modelling of information and its processing paved the way for a host of new areas such as quantum computing, quantum communication, and quantum cryptography. An introduction to these concepts and an overview of scientific case studies is summarized in\cite{alexeev2019quantum}. The development of quantum computing was motivated by the idea that certain quantum effects could possibly be used to speed-up computing beyond what is possible in the classical realm. Quantum computing differs from its classical counterpart in that the basic unit of computation is a qubit, which can take a continuum of values, thereby allowing storage and processing of quantum superpositions of data. The non-classical way of encoding the information using qubits enables quantum algorithm to benefit from quantum phenomena such as quantum annealing, quantum tunneling, and quantum entanglement. 
Quantum algorithm defines the logic used in the processing of quantum states, whereas the implementation is characterized by the chosen quantum computational mode. The quantum computational modes differ in terms of tunability and control. Examples include universal gate, quantum annealing, and adiabatic quantum computing.  Moreover, the structure of the algorithm may have implications on the selection of optimal mode. The universal gate mode is generally considered to be the most powerful, and it resembles the universal gate mode used in classical computing. A key limitation of universal gate model is that its technology is not mature and currently supports only a few qubits. Certain non-universal modes such as adiabatic quantum computing can be made universal in some cases. On the other hand, quantum annealers are most advanced from a hardware perspective and latest implementations consist of around 2000 qubits. A detailed overview of quantum annealing and applications in quantum computing is given in\cite{biswas2017nasa}. In the rest of this section, we discuss this in more detail and explain different aspects of applying this approach to optimization problems relevant to mobile network automation. 

\subsection{Quantum Annealer}
Quantum annealer computation mode is based on optimization of energy function using quantum fluctuations. The input problem is mapped to the classical Ising-spin problem, which leads to a general framework for analog quantum computation. Apart from a large number of qubits, a key reason of its relevance to the field of computer science is that it provides a generalized framework to solve combinatorial optimization problems. In order to increase it accessibility, low-level details have been abstracted away. As a consequence, deep knowledge of quantum mechanics is not needed for programming the solver. The quantum annealer based computational device implements quantum annealing, which is a general metaheuristic optimization algorithm that searches the solution space efficiently by using quantum effects such as quantum tunneling. The underlying idea is similar to classical local search algorithms such as simulated annealing, which makes it accessible to users not well versed in quantum mechanics. Simulated
annealing makes use of thermal fluctuation to overcome energy barriers and avoid entrapment in local minima. In contrast, quantum annealing makes use of quantum tunneling.

Quantum annealer based quantum processing units~(QPUs) are designed to run quantum annealing algorithm to solve QUBO problems. Therefore, the first step towards solving a SON use-case computational problem using quantum annealer is to formulate it as a QUBO. As mentioned previously, a wide class of optimization problems of practical nature, pertaining to RAN optimization use-cases, can be expressed as cost functions over discrete sets of binary variables. Reformulating these problems as optimization of quadratic functions over
binary variables paves the way for their solution via a quantum computing based mobile network automation framework discussed in the rest of this section.
\begin{figure}[t]
\begin{center}
\includegraphics[trim={0.35cm 0cm 0cm 0cm},scale=0.30,clip]{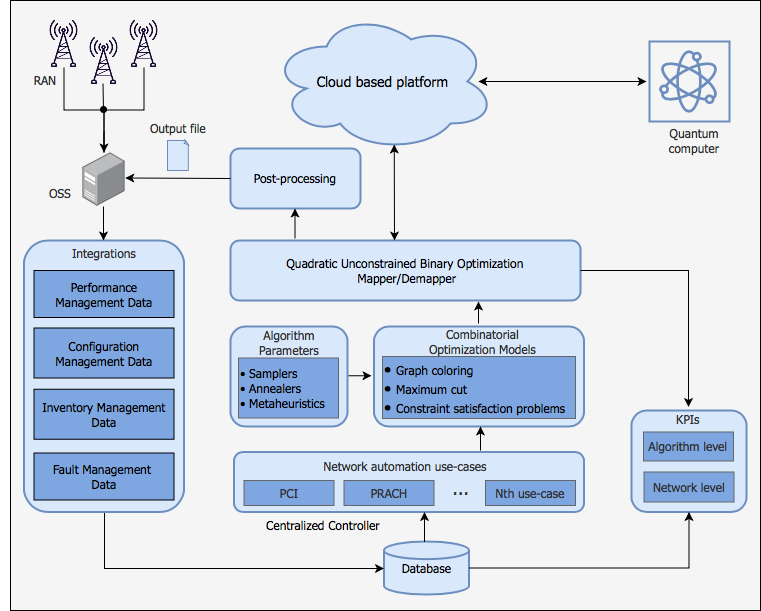}
\end{center}
\caption{A QUBO based cloud quantum computing framework for mobile network automation.}
\vspace{-7mm}
\label{fig:framework}
\end{figure}
\subsection{Network Automation Framework}
We consider a centralized SON (CSON) architecture consisting of a cloud based quantum annealer for solving
computational problems for SON use-cases. The system processes
aggregated data and outputs a plan to update configuration
parameters in the whole network. In order to support a wide range of SON use-cases, different types
of data such as configuration management, performance
management, inventory management, and fault management is fetched
from the network and stored in the environment.
The configuration management data consists of all the configuration parameters
of the RAN,
frequency bands and channels, neighbor relations
between cells etc. On the other
hand, performance management data consists of various performance counters collected
by the RAN. Compared to configuration management data, it is updated very
frequently. The counter data is in the form of time-series and
typically has a very high granularity, e.g. of the order of fifteen
to thirty minutes. In addition cell site and antenna related
information such as geographical location, antenna type, and
bearings is available in the inventory management data. Finally, fault management data consists of network alarms and is useful for SON use-cases meant for the fault diagnosis and self-healing. 

As illustrated in Fig.~\ref{fig:framework}, implementation of a SON use-case using the proposed framework, requires a QUBO model of the underlying computational problem. Note that creation of combinatorial optimization models and the subsequent selection of algorithm and related optimization parameters for each use-case, needs to be done by the data scientists in collaboration with the network optimization experts. The three models mentioned here (graph coloring, maximum-cut, and CSPs) are commonly used for network optimization problems. All these can be converted to QUBO form~\cite{kochenberger2004unified}. To this end, SDKs and other tools supported by the quantum computing environment may be used. Moreover, for new use-cases, the accuracy of the model and effectiveness of the algorithm approach can be verified at a small scale during this step. The model for the full network is then submitted to the quantum computing cloud, which runs the quantum annealer and returns the solution. The post-processing phase processes the solution to create network plans with modified parameters, to be provisioned to the network through the vendor operation support system~(OSS). Moreover, key performance indicators~(KPIs) are collected from both the network and the computing environment to track and analyze the performance of the network as well as the algorithm. In the following sections, we discuss a case study on PRACH optimization using the proposed framework, followed by an empirical analysis of the performance of classic heuristics and quantum approaches. 

\section{A Case Study On PRACH RSI Assignment}

\subsection{RSI assignment in LTE/NR Radio Access Networks}
In LTE/LTE-A, random access procedure is used to acquire uplink synchronization and access the network for transmitting signaling and data. It involves transmission of a preamble by the user to eNB.
%There are two main types of LTE random access, namely, contention-based and contention-free. In contention-based procedure, UE randomly selects a preamble from the set of preambles and transmits it to the eNB. It may lead to a collision where more than one UEs simultaneously transmit their respective preambles. In such situation, collision resolution procedure is followed, resulting in unpredictable latency.  Contention-free procedure simply alleviates this problem by allocating a dedicated preamble to UE for random access. Contention-free access is faster, and is useful for use-cases that require low latency such as handovers and uplink synchronisation for downlink data transfer in RRC CONNECTED state.
In order to maximize the orthogonality between users performing random access channel procedure, Zadoff-Chu~(ZC) sequences are used to generate preambles in each cell. The ZC sequences are used due to their constant amplitude and good auto/cross-correlation properties. 
%In particular, low cross-correlation between different ZC sequences, and zero auto-correlation between cyclic-shifted version of the same sequence.
%This preamble is selected from a set of 64 preambles, which are cyclically shifted versions of Zadoff-Chu root sequences. The number of root sequences needed to generate 64 preambles depends on a multitude of factors such as number of possible cyclic shifts and the speed of UEs. 
Total number of ZC root sequences according to 3GPP is $838$. Therefore, reuse is inevitable and proper planning is needed. %Each root sequence is 839 bits in length and spans 800 ms.  
Depending on cell radius, more than one root sequence may be required. A user needs to know which sequences it is allowed to use to generate the require preambles. This information needs to be signalled in the cell in an efficient manner. To this end, the index of only the first sequence, i.e. RSI, is broadcasted in the cell. The order of sequences is predefined, on the basis of criteria such as configuration properties, maximum number of cyclic shifts. Therefore, it possible for user to derive the preambles using RSI as a base index. Thus, RSI is a logical parameter that is used as a base to calculate the preambles. It is worth mentioning that this logical index is mapped to a physical RSI during the implementation phase.
The user gets information about the cyclic shift from the zero correlation zone configuration and applies it to the base index to generate the preambles. Resulting preamble sequences should not overlap with the sequences in neighbour cells. Overlapping preamble sequences lead to reservation of physical downlink control channel~(PDCCH) and physical uplink shared channel~(PUSCH) resources in cells that receive such ghost preambles. Due to the limited number of RSIs, its not possible to assign a unique RSI to every cell in the network. Under such practical constraints, reuse distance is defined and assignment of RSIs is done in a way that RSIs within the re-use cluster do not conflict. Thus, assignment of non-conflicting RSIs to cell is critical to the generation of correct preambles. Moreover, set of RSIs for a given cell radius can be calculated by using the aforementioned approach. It is worth noting that in the graph coloring model addressed in the current paper, set of RSIs represents the set of colours available to colour the network graph. 
\subsection{Network Graph Model}
We model LTE network by an undirected graph $G(\mathcal{V},\mathcal{E})$ comprising a set $\mathcal{V}$ of vertices and a set $\mathcal{E}$ of edges. An edge $e_{i,j} \in \mathcal{E}$  connects a pair of vertices in $v_i, v_j \in \mathcal{V}$, for  $i\ne j$. Two vertices connected by an edge are said to be adjacent to each other. Set of vertices adjacent to the vertex $v_i$ constitutes neighborhood $\mathcal{N}_{v_i}$, and the number of edges incident to vertex $v_i$ is known as the degree of $v_i$ denoted by deg($v_i$). The vertices and edges denote LTE cells in the network, and the neighbor relations that exist between them, respectively. Neighbour relations or adjacencies are based on mutual interference between cells, which is dependent on a number of propagation factors including spatial separation and antenna directions. According to graph coloring terminology, a $N$-coloring for a given set of colors $\mathcal{N} = \left\{ 1,\dots,N\right\}$, is defined as function  $c: \mathcal{V} \rightarrow \mathcal{N}$. Two vertices $v_i, v_j$ connected by an edge are said to be conflicting if they are assigned the same color. %, i.e., $c(v_i)=c(v_j)$. 
In contrast, $N$-coloring is defined to be \emph{legal}, if colors are assigned to the vertices in a way that there are no conflicting edges. Here, we are interested in legal $N$-coloring of $G(\mathcal{V},\mathcal{E})$, where $N = |\mathcal{N}|$ is the cardinality of the set of valid RSIs. Next, we discuss some classic heuristics for solving this graph coloring problem. These results will be used as a baseline for analyzing the performance of the quantum computing enabled framework.

\subsection{Classic Heuristics Approach}
The NP-hardness of the graph coloring problem necessitates the use of fast heuristics for the determination of a (suboptimal) solution, especially in practical scenarios where the underlying graph is dynamic and mandates the use of polynomial time algorithms. We consider heuristics based on the greedy principle --- a highly intuitive approach where a locally optimal solution is picked at every decision step. In the context of graph coloring, greedy algorithm refers to the approach where colors are assigned to vertices in a direct sequential manner. Resulting algorithms are suboptimal but fast, making it possible to find a solution in polynomial time. This approach is also called sequential coloring as it colors the vertices in a given sequence.
Input arguments include a graph (i.e. sets of vertices and edges), and sequence $S = ( v_1, \dots, v_I )$ which is used by the algorithm to assign the colors. Design of $S$ is critical to the performance of algorithm and can be done in a number of ways. In what follows, we briefly discuss a few greedy algorithms.
\subsubsection{Random Sequential (RS)}
A basic variant of sequential coloring can be devised by simply using a random sequence of vertices $S = ( v_1, \dots, v_N )$. Accordingly, it is called random sequential, and is considered as a baseline for evaluating the performance of sequential coloring algorithms on a given class of graphs. It is relatively simpler to implement as it does not entail any logic for sequence generation. 

\subsubsection{Connected Sequential (CS)}
A sequential coloring is called connected sequential coloring if the colored vertices induce a connected graph. The underlying principle is that during each iteration of the coloring algorithm, only the vertices adjacent to the ones already colored are considered as candidates for coloring. In many cases, this modification helps in avoiding local optima, which leads to an optimal coloring. 

\subsubsection{Independent Set (IS)}
Independent set is based on the maximal independent set algorithm. In the first step, a maximal set of vertices is computed. All vertices in the set are assigned the first available color. Colored vertices are subsequently removed from the graph and the same procedure is repeated. For example, consider color $c_i$, assigned to all possible vertices in a maximal independent set (an important property of such a set is that no two vertices are adjacent). In the next step, the algorithm removes all these vertices from the original set of vertices, and continues with the remaining subgraph and color $c_{i+1}$. Note that the maximal independent set computed in each step ensures that the maximum number of vertices are assigned the next available color.

\subsubsection{Largest First (LF)}
An effective way to create a sequence of vertices for sequential coloring is to sort them according to vertex degree in a descending order. The rationale behind this highly intuitive approach is that vertices with higher degree are harder to color and impose more constraint on the number of colors, therefore it makes sense to color them first. 

\subsubsection{DSATUR (DS)}
The DSATUR also known as saturation last first algorithm makes use of similar underlying principle as largest first algorithm, i.e. color harder vertices first. However, it takes into account degree of saturation $\rho(v)$ (hence the name DSATUR), which is defined as the number of distinct colors used in the neighborhood of vertex $v$. It works because constraints on the set of available colors for a given vertex $v$ are indeed dependent on the set of colors used in the neighborhood rather than the degree deg$(v)$. 

\subsection{Quantum Computing Approach}
The first step towards solving a combinatorial optimization problem using quantum computing involves mathematical formulation supported by the underlying computer architecture. To this end, we consider QUBO formulation, which has recently been proposed as it can be used to model a number of combinatorial optimization problems including graph coloring. We use DWAVE-Leap cloud based quantum computing platform that supports QUBO problems~\cite{dwave_leap}. In order to formulate a QUBO, we note that graph coloring is inherently a CSP, where the objective is to find a configuration or an assignment of colors that doesn't violate the constraints. In this case, there are two constraints:
\begin{itemize}
    \item Each vertex is assigned only one color.
    \item No two vertices connected by an edge have the same color.
\end{itemize}

It is known that the graph coloring CSP with the above constraints can be formulated as a QUBO problem involving minimization of $\mathbf{x}^{T} \mathbf{Q} \mathbf{x}$,
%\begin{equation}
%\begin{array}{cl}
%\hspace{0mm}\text{minimize} \hspace{3mm} \mathbf{x}^{T} \mathbf{Q} \mathbf{x}
%\end{array} \nonumber 
%\label{eq:2.12}
%\end{equation}
where $\mathbf{x}^T = [x_1, \dots x_ {m}, \dots x_M]$ is a vector of binary decision variables (i.e. $x_m \in \mathcal{\{0, 1\}}$) that reflects qubits, and $\mathbf{Q} \in \mathbb{R}^{M \times M}$ is matrix comprising of weights corresponding to the penalties for the violation of constraints. Binary representation is essential because the solution is implemented using qubits with spins. Also, in this formulation, constraints are embedded in the objective function. Diagonal elements of $\mathbf{Q}$ can be understood as qubit weights or biases, whereas off-diagonal elements are inter-qubit couplings. The elements are chosen such that minimization of objective function leads to the desired ground state where the constraints are met. Note that $M = N \times K$ is the number of optimization variables or logical qubits. Once the QUBO problem is solved, these can be transformed back to reveal the color assignment of the $N$ vertices. Next, the logical qubits are mapped to the physical qubits and couplers, on the QPU chimera graph architecture. It is pertinent to note that the problem size is limited by the number of available qubits. Problems with variables exceeding the qubits require an additional step of decomposition into smaller sub-problems. These are solved using hybrid approaches, comprising of a combination of metaheuristics and quantum annealing. The QUBO formulation is submitted to the DWAVE solver, which returns a sample of solutions along with the corresponding energy levels. If the problem instance is solved successfully, the solution with the lowest energy level is the best possible solution, and it meets all the constraints considered in the QUBO formulation.

%Let us express the previously formulated graph coloring problem as a binary CSP, with variables given by the set $\mathcal{V}$, where each variable takes a value from domain $\mathcal{N}$. In order to formalize the constraints, we introduce binary variables. Let $x_{i, n} = 1$ if $c(v_i) = n$ and zero otherwise. Then, network-wide constraint $\mathcal{C}_1$ can be expressed as
%\begin{equation}
%\mathcal{C}_1: \quad \sum_{i \in \mathcal{I}}\sum_{n \in \mathcal{N}} x_{i,n} = %1
%\label{eq:2.1}
%\end{equation}
%This constraints reflects that each vertex is assigned only one color. Next, we formulate constraint $\mathcal{C}_2$ comprising binary variables $x_{i,n}$ and $x_{j,n}$ to ensure that vertices $v_i$ and $v_j$ connected by an edge $e_{i,j}$ are not assigned the same color.
%\begin{equation}
%\mathcal{C}_2: \quad x_{i,n} + x_{j,n} = 1 \quad \forall~n \in \mathcal{N}.
%\label{eq:2.1}
%\end{equation}
%These constraints are subsequently used to formulate a QUBO problem for quantum annealing solver as discussed in Section~\Rmnum{4}. Draw two arch diagrams here. Data->Network Graph->Coloring. Data-->Network Graph> CSP>QUBO>Coloring

\subsection{Results}
In order to construct a graph for comparing classical and quantum computing approaches, we use data from a cluster of cells in a commercial LTE network. Apart from relevant configuration parameters such as RSI and frequency channel number, it includes geographical locations of cells and antenna bearings. Antenna alignment and inter-cell distance is used to calculate cost of RSI conflict between a pair of cells. For a given source cell, list of geographical neighbor cells within a certain radius and the same frequency channel are ranked on the basis of cost. The cells with rank less than or equal to \emph{conflict rank} are picked as candidate neighboring cells to avoid RSI conflicts. Recall that cell is represented by a vertex in the graph. Therefore, vertex corresponding to each cell in the candidate neighboring cells is connected to the source cell vertex via an edge. This procedure is run by all the cells, which leads to an undirected graph with coloring difficulty parameterized by \emph{conflict rank}. Coloring the graph with increased conflict rank leads to conflict free RSI assignment with higher reuse distance. In real networks, this parameter is fixed to a certain value that ensures PRACH related KPI targets are met. Here, it is used to increase the complexity of the RSI assignment problem in order to analyze various coloring approaches.

\begin{figure}[tbh]
\begin{center}
\includegraphics[trim={0.25cm 0cm 0cm 0.1cm},scale=0.3,clip]{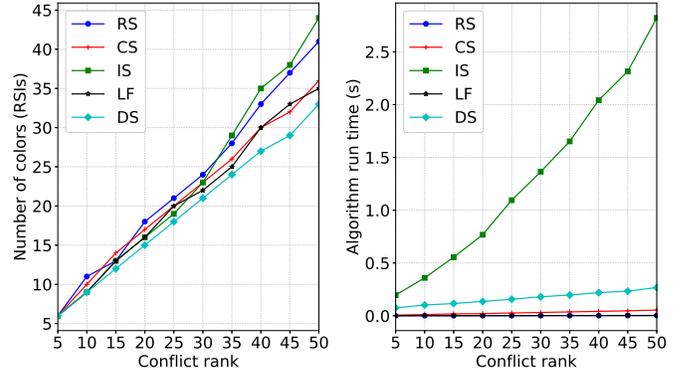}
\end{center}
\vspace{-5mm}
\caption{Performance comparison of different classic algorithms in terms of number of RSIs required for conflict-free assignment (left) and corresponding algorithm run time (right).}
\label{fig:classic}
\end{figure}

\begin{figure}[tbh]
\begin{center}
\includegraphics[trim={0.25cm 0cm 0cm 0.1cm},scale=0.3,clip]{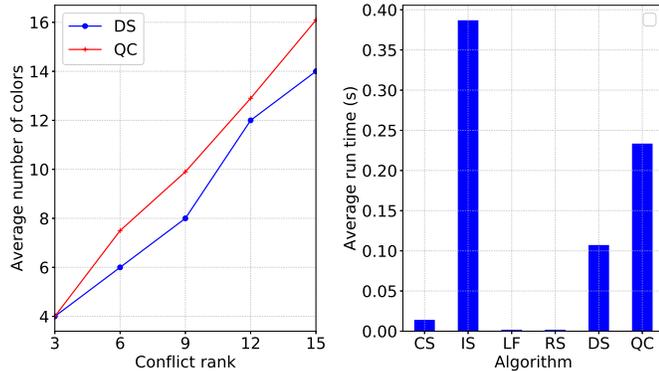}
\end{center}
\vspace{-5mm}
\caption{Performance comparison of quantum computing approach with classic DSATUR~(DS) algorithm (left), and average run times (right).}
\label{fig:quantum}
\end{figure}

First, we analyze the performance of coloring approaches in the context of classical computing discussed in Section~\Rmnum{3}~C. These include random sequential (RS), independent set (IS), connected sequential (CS), least first (LF), and DSATUR (DS). The algorithms are run to the color the network graph, and performance is analyzed in terms of number of colors used by the algorithm and the corresponding average run time of the algorithm. In order to study the impact of increased problem complexity on performance, network graphs are generated for a range of conflict ranks. Results are averaged over a large number of runs, for a graph instance corresponding to a given conflict rank. Figure~\ref{fig:classic} shows that difference in performance becomes evident as the conflict rank increases. In terms of number of colors, DS outperforms the rest, followed by CS and LF, whereas RS and IS clearly need more colors. 
However, in terms of algorithm run time, LF and RS perform the best, followed by CS, DS, and finally IS. The trends observed in run time can be explained by the big-O analysis of the algorithms. Next, we turn our attention to quantum computing approaches. The aim is to color the same network graphs using a quantum computer and compare results to the classic coloring algorithms running on a regular laptop computer. To this end, DWAVE systems' Leap quantum cloud service is used~\cite{dwave_leap}, which provides real-time access to DWAVE 2000Q QPU consisting of more than 2000 qubits. We use the DWAVE software tools including the cloud based pre-built integrated development environment~(IDE) and the python SDK to develop code for solving the graph coloring problem. The code is essentially a python script that builds the graph coloring problem and calls the vertex coloring function, which defines a QUBO with ground states correspond to minimum vertex coloring. A sampler is specified to sample from the low energy states of the defined QUBO. It returns an iterable of samples in the order of increasing energy. Due to the large size of the problem, we choose the built-in Kerberos sampler that is hybrid in that it runs three branches in parallel namely tabu search, simulated annealing, and QPU. The QPU is used to solve the subproblem comprising of high impact problem variables. Related input parameters are set to recommended values. These include maximum number of iterations max\_iters = $100$, iterations with no improvement that terminate sampling convergence\_iters = $3$, and maximum size of subproblem selected in the QPU branch max\_subproblem\_size = $50$. Upon successful termination, the best sample is the solution as it corresponds to the lowest energy state. 
When the code is run in the IDE, the problem is submitted to the quantum computing platform, which solves it and returns the solution. In addition to the outcome and the status of the solver, detailed results can be queried using SDK. Moreover, solver properties and performance statistics such as time elapsed during the computation can also be accessed via an online dashboard. This makes it possible to know the time taken by the platform to solve the coloring problem. 
The results shown in Fig.~\ref{fig:quantum} show that quantum computing~(QC) approach can successfully color the graph for a range of conflict ranks. 
We compare the results obtained from QC solver to DS, which is the best performing coloring algorithm from a classic computing platform. It is clear that solution quality of QC solution falls short of DS, i.e. DS is able to color the graph with less number of colors. In terms of computation time aggregated over conflict ranks and averaged over multiple runs, DS again outperforms QC by an order of magnitude. Apart from IS, rest of the algorithms perform even better, which is in line with the run time results in Fig.~\ref{fig:classic}. It is clear that in this case, even though quantum approach successfully solves the RSI assignment problem, it offers no clear benefit over classic graph coloring approaches. In particular, there is no quantum speedup observed. It is important to note that this result is not surprising, as for most problems classic algorithms outperform current quantum computing solutions~\cite{mandra2018deceptive}. This is due to a number of factors: first, the implementation is a semi-quantum algorithm, which is a necessity in this case because of large problem size, and second, the results are dependent on the type of quantum computer. In general, for non-trivial problems, gains are theoretically possible, but require extremely high number of qubits, and often gains vanish once the overhead are taken into account~\cite{campbell2019applying}. Nevertheless, the quantum annealing based approach makes use of the QUBO model, which enables a powerful and flexible framework that can be used for solving a wide range of network optimization problems.

\section{Conclusion}
We have discussed the application of quantum computing approach for mobile network optimization. To this end, formulation of network optimization as a CSP is discussed. The problems are mapped to a QUBO problem, subsequently solved using a cloud based commercial quantum computer. As a case study, we discuss PRACH RSI assignment problem, which is relevant to both 4G and 5G networks. Results show that the QUBO based quantum computer can successfully solve the RSI optimization problem for a range of complexity levels. 
However, it is worth mentioning that the current implementation is semi-quantum, and number of qubits are limited, thereby leading to no quantum speed-up. This is inline with a number of recent studies that show that for most practical problems of interest, quantum computing technology is currently not mature enough to realize the theoretical gains. We conclude that due to its generic nature and the capability to solve a wide range of network optimization problems formulated as CSPs, the presented QUBO based framework is fundamentally well-suited for harnessing the power of quantum computing for mobile network optimization and will become increasingly useful as the quantum computing technology progresses.

%\ifCLASSOPTIONcaptionsoff
%  \newpage
%\fi

\bibliographystyle{IEEEtran}
\bibliography{SON}

% Generated by IEEEtran.bst, version: 1.14 (2015/08/26)
\begin{thebibliography}{10}
\providecommand{\url}[1]{#1}
\csname url@samestyle\endcsname
\providecommand{\newblock}{\relax}
\providecommand{\bibinfo}[2]{#2}
\providecommand{\BIBentrySTDinterwordspacing}{\spaceskip=0pt\relax}
\providecommand{\BIBentryALTinterwordstretchfactor}{4}
\providecommand{\BIBentryALTinterwordspacing}{\spaceskip=\fontdimen2\font plus
\BIBentryALTinterwordstretchfactor\fontdimen3\font minus
  \fontdimen4\font\relax}
\providecommand{\BIBforeignlanguage}[2]{{%
\expandafter\ifx\csname l@#1\endcsname\relax
\typeout{** WARNING: IEEEtran.bst: No hyphenation pattern has been}%
\typeout{** loaded for the language `#1'. Using the pattern for}%
\typeout{** the default language instead.}%
\else
\language=\csname l@#1\endcsname
\fi
#2}}
\providecommand{\BIBdecl}{\relax}
\BIBdecl

\bibitem{neukart2017traffic}
F.~Neukart, G.~Compostella, C.~Seidel, D.~Von~Dollen, S.~Yarkoni, and
  B.~Parney, ``Traffic flow optimization using a quantum annealer,''
  \emph{Frontiers in ICT}, vol.~4, p.~29, 2017.

\bibitem{venturelli2019reverse}
D.~Venturelli and A.~Kondratyev, ``Reverse quantum annealing approach to
  portfolio optimization problems,'' \emph{Quantum Machine Intelligence},
  vol.~1, no. 1-2, pp. 17--30, 2019.

\bibitem{ORUS2019100028}
R.~Or{\'u}s, S.~Mugel, and E.~Lizaso, ``Quantum computing for finance: Overview
  and prospects,'' \emph{Reviews in Physics}, vol.~4, p. 100028, 2019.

\bibitem{Biamonte2017}
J.~Biamonte, P.~Wittek, N.~Pancotti, P.~Rebentrost, N.~Wiebe, and S.~Lloyd,
  ``Quantum machine learning,'' \emph{Nature}, vol. 549, no. 7671, pp.
  195--202, 2017.

\bibitem{qml_6g}
S.~J. {Nawaz}, S.~K. {Sharma}, S.~{Wyne}, M.~N. {Patwary}, and
  M.~{Asaduzzaman}, ``Quantum machine learning for 6{G} communication networks:
  State-of-the-art and vision for the future,'' \emph{IEEE Access}, vol.~7, pp.
  46\,317--46\,350, 2019.

\bibitem{MITCHELL2020101825}
C.~J. Mitchell, ``The impact of quantum computing on real-world security: A
  5{G} case study,'' \emph{Computers \& Security}, vol.~93, p. 101825, 2020.

\bibitem{challengesSON}
{A. Imran, A. Zoha, A. Abu-Dayya}, ``{Challenges in 5G: How to Empower {SON}
  with Big Data for Enabling 5G},'' \emph{IEEE Network}, vol.~28, no.~6, pp.
  27--33, 2014.

\bibitem{kochenberger2004unified}
G.~A. Kochenberger, F.~Glover, B.~Alidaee, and C.~Rego, ``A unified modeling
  and solution framework for combinatorial optimization problems,'' \emph{OR
  Spectrum}, vol.~26, no.~2, pp. 237--250, 2004.

\bibitem{Hamiltonpostmoore}
\BIBentryALTinterwordspacing
K.~E. Hamilton, C.~D. Schuman, S.~R. Young, R.~S. Bennink, N.~Imam, and T.~S.
  Humble, ``Accelerating scientific computing in the post-moore’s era,''
  \emph{ACM Trans. Parallel Comput.}, vol.~7, no.~1, Mar. 2020. [Online].
  Available: \url{https://doi.org/10.1145/3380940}
\BIBentrySTDinterwordspacing

\bibitem{LTE_JECE}
F.~Ahmed, O.~Tirkkonen, M.~Peltom\"aki, J.-M. Koljonen, C.-H. Yu, and M.~Alava,
  ``Distributed graph coloring for self-organization in \uppercase{LTE}
  networks,'' \emph{J. Elec. Computer Engineering}, pp. 1--10, 2010.

\bibitem{alexeev2019quantum}
Y.~Alexeev, D.~Bacon, K.~R. Brown, R.~Calderbank, L.~D. Carr, F.~T. Chong,
  B.~DeMarco, D.~Englund, E.~Farhi, B.~Fefferman \emph{et~al.}, ``Quantum
  computer systems for scientific discovery,'' \emph{arXiv preprint
  arXiv:1912.07577}, 2019.

\bibitem{biswas2017nasa}
R.~Biswas, Z.~Jiang, K.~Kechezhi, S.~Knysh, S.~Mandra, B.~O’Gorman,
  A.~Perdomo-Ortiz, A.~Petukhov, J.~Realpe-G{\'o}mez, E.~Rieffel \emph{et~al.},
  ``A nasa perspective on quantum computing: Opportunities and challenges,''
  \emph{Parallel Computing}, vol.~64, pp. 81--98, 2017.

\bibitem{dwave_leap}
D-Wave, ``D-wave system documentation,'' Available at
  \url{https://docs.dwavesys.com/docs/latest/index.html} (2021/01/30).

\bibitem{mandra2018deceptive}
S.~Mandra and H.~G. Katzgraber, ``A deceptive step towards quantum speedup
  detection,'' \emph{Quantum Science and Technology}, vol.~3, no.~4, p. 04LT01,
  2018.

\bibitem{campbell2019applying}
E.~Campbell, A.~Khurana, and A.~Montanaro, ``Applying quantum algorithms to
  constraint satisfaction problems,'' \emph{Quantum}, vol.~3, p. 167, 2019.

\end{thebibliography}

% biography section
% 
% If you have an EPS/PDF photo (graphicx package needed) extra braces are
% needed around the contents of the optional argument to biography to prevent
% the LaTeX parser from getting confused when it sees the complicated
% \includegraphics command within an optional argument. (You could create
% your own custom macro containing the \includegraphics command to make things
% simpler here.)
%\begin{IEEEbiography}[{\includegraphics[width=1in,height=1.25in,clip,keepaspectratio]{mshell}}]{Michael Shell}
% or if you just want to reserve a space for a photo:

% if you will not have a photo at all:
%\begin{IEEEbiographynophoto}{John Doe}
%Biography text here.
%\end{IEEEbiographynophoto}

% insert where needed to balance the two columns on the last page with
% biographies
%\newpage

%\begin{IEEEbiographynophoto}{Furqan Ahmed}
%Biography text here.
%\end{IEEEbiographynophoto}

% You can push biographies down or up by placing
% a \vfill before or after them. The appropriate
% use of \vfill depends on what kind of text is
% on the last page and whether or not the columns
% are being equalized.

%\vfill

% Can be used to pull up biographies so that the bottom of the last one
% is flush with the other column.
%\enlargethispage{-5in}

% that's all folks
\end{document}